\documentclass[preprintnumbers,amsmath,amssymbm,prd]{revtex4}
\usepackage{epsfig}
\usepackage{graphicx}
\usepackage{amssymb}

\begin{document}
\title{No nonminimally coupled massless scalar hair for spherically symmetric neutral reflecting stars}
\author{Shahar Hod}
\affiliation{The Ruppin Academic Center, Emeq Hefer 40250, Israel}
\affiliation{ }
\affiliation{The Hadassah Institute, Jerusalem 91010, Israel}
\date{\today}

\begin{abstract}
\ \ \ It has recently been proved that horizonless compact stars
with reflecting boundary conditions {\it cannot} support spatially
regular matter configurations made of minimally coupled scalar
fields, vector fields, and tensor fields. In the present paper we
extend this intriguing no-hair property to the physically
interesting regime of scalar fields with {\it nonminimal} coupling
to gravity. In particular, we prove that static spherically
symmetric configurations made of nonminimally coupled massless
scalar fields cannot be supported by compact reflecting stars.
\end{abstract}
\bigskip
\maketitle

\section{Introduction}

The early no-scalar-hair theorems of Chase \cite{Chas}, Bekenstein \cite{Bek1}, and
Teitelboim \cite{Teit} have rigorously proved that asymptotically flat black holes with classical absorbing horizons
cannot support nontrivial external configurations made of
static massless or massive scalar fields {\it minimally} coupled to gravity \cite{Noteit,Hodrc,Herkr}.
Later no-scalar-hair theorems \cite{Heu,Bek2,BekMay,Bek20} have intriguingly ruled out the
existence of spherically symmetric static hairy black-hole configurations made of
minimally coupled scalar fields with positive semidefinite self-interaction potentials.

For the mathematically more challenging case of scalar fields with
{\it nonminimal} coupling to gravity, one may use the elegant
no-hair theorems of Bekenstein and Mayo \cite{BekMay,Bek20,Notesa}
to rigorously exclude, in the physical regimes $\xi<0$ and
$\xi\geq1/2$ of the dimensionless coupling parameter \cite{Notewx},
the existence of spherically symmetric asymptotically flat black
holes with classical absorbing horizons that support static
spatially regular scalar fields with positive semidefinite
self-interaction potentials. Recently, a novel no-scalar-hair
theorem has been derived which rigorously rules out the existence of
spherically symmetric static hairy black-hole configurations made of
massless scalar fields for {\it generic} values of the physical
coupling parameter $\xi$ \cite{Hodn}.

The mathematically elegant and physically interesting no-scalar-hair
theorems presented in \cite{Chas,Bek1,Teit,Heu,Bek2,BekMay,Bek20},
which rigorously exclude the existence of asymptotically flat static
composed black-hole-scalar-field hairy configurations with absorbing
event horizons, naturally motivate the following physically
intriguing question: can the characteristic no-scalar-hair behavior
of static black-hole spacetimes
\cite{Chas,Bek1,Teit,Heu,Bek2,BekMay,Bek20} be extended to the
physical regime of {\it horizonless} curved spacetimes?

In order to address this physically interesting question, we have
recently \cite{Hodns} explored the static sector of the non-linearly
coupled Einstein-scalar field equations with reflecting ({\it
repulsive}) boundary conditions at the surface of a spherically
symmetric {\it horizonless} compact star. Interestingly, it has been
explicitly proved in \cite{Hodns} that horizonless compact objects
with reflecting boundary conditions (compact reflecting stars
\cite{Noteref}) share the intriguing no-scalar-hair property with
the more familiar classical black-hole spacetimes with absorbing
({\it attractive}) event horizons. In particular, we have revealed
the fact \cite{Hodns} that spherically symmetric neutral reflecting
stars cannot support non-linear static scalar (spin-0) fields {\it
minimally} coupled ($\xi=0$) to gravity. The no-hair behavior
observed in \cite{Hodns} was later extended in the interesting work
of Bhattacharjee and Sarkar \cite{Bha}, who have proved that compact
objects (horizonless reflecting stars) cannot support higher-spin
[vector (spin-1) and tensor (spin-2)] fields.

The main goal of the present paper is to extend the no-hair behavior
recently observed in \cite{Hodns,Bha} for horizonless compact
objects to the physical regime of non-linear scalar fields with {\it
nonminimal} coupling to gravity. Interestingly, exploring the
non-linearly coupled Einstein-scalar field equations, we shall
explicitly prove below that spherically symmetric horizonless
compact reflecting stars, like the more familiar absorbing
black-hole spacetimes, cannot support static nonminimally coupled
massless scalar fields with {\it generic} values of the
dimensionless coupling parameter $\xi$.

\section{Description of the system}

We consider a spherically symmetric neutral reflecting star
\cite{Noteref} of radius $R_{\text{s}}$ which interacts non-linearly
with a massless scalar field $\psi$. We shall assume that the scalar
field has a nonminimal coupling to the characteristic scalar
curvature $R$ of the spacetime [see Eq. (\ref{Eq3}) below]. The
static and spherically symmetric composed star-field configurations
can be described by the curved line element \cite{BekMay,Noteunit}
\begin{equation}\label{Eq1}
ds^2=-e^{\nu}dt^2+e^{\lambda}dr^2+r^2(d\theta^2+\sin^2\theta
d\phi^2)\ ,
\end{equation}
where $\nu=\nu(r)$ and $\lambda=\lambda(r)$. We consider
asymptotically flat spacetimes which are characterized by the simple
asymptotic functional relations \cite{BekMay}
\begin{equation}\label{Eq2}
\nu\sim M/r\ \ \ \text{and}\ \ \ \lambda\sim M/r\ \ \ \ \text{for}\
\ \ \ r\to\infty\  ,
\end{equation}
where $M$ is the total (asymptotically measured) mass of the static
spacetime.

The non-minimally coupled massless scalar field $\psi$ is
characterized by the action \cite{BekMay,NoteSEH}
\begin{equation}\label{Eq3}
S=S_{EH}-{1\over2}\int\big(\partial_{\alpha}\psi\partial^{\alpha}\psi+\xi
R\psi^2\big)\sqrt{-g}d^4x\ ,
\end{equation}
where the dimensionless physical parameter $\xi$ quantifies the
strength of the nonminimal coupling of the field to the Ricci scalar
curvature $R(r)$ of the spherically symmetric spacetime. The action
(\ref{Eq3}) yields the functional expressions \cite{BekMay,Notetag}
\begin{equation}\label{Eq4}
T^{t}_{t}=e^{-\lambda}{{\xi(4/r-\lambda{'})\psi\psi{'}+(2\xi-1/2)(\psi{'})^2+2\xi\psi\psi{''}}
\over{1-8\pi\xi\psi^2}}\ ,
\end{equation}
\begin{equation}\label{Eq5}
T^{r}_{r}=e^{-\lambda}{{(\psi{'})^2/2+\xi(4/r+\nu{'})\psi\psi{'}}
\over{1-8\pi\xi\psi^2}}\  ,
\end{equation}
and
\begin{equation}\label{Eq6}
T^{t}_{t}-T^{\phi}_{\phi}=e^{-\lambda}{{\xi(2/r-\nu{'})\psi\psi{'}}
\over{1-8\pi\xi\psi^2}}
\end{equation}
for the components of the energy-momentum tensor which characterizes the nonminimally
coupled massless scalar field. As explicitly proved in
\cite{BekMay}, the mixed components of the energy-momentum tensor
must be finite for generic physically acceptable systems:
\begin{equation}\label{Eq7}
\{|T^{t}_{t}|,|T^{r}_{r}|,|T^{\theta}_{\theta}|,|T^{\phi}_{\phi}|\}<\infty\
.
\end{equation}
In particular, for asymptotically flat spacetimes with finite total mass
(as measured by asymptotic observers), the energy density
$\rho\equiv -T^{t}_{t}$ of the matter fields is characterized by the
asymptotic functional behavior \cite{Hodpb}:
\begin{equation}\label{Eq8}
r^3\rho(r)\to0\ \ \ \text{for}\ \ \ r\to\infty\  .
\end{equation}

In addition, as explicitly proved in \cite{BekMay}, causality
requirements enforce the simple relations
\begin{equation}\label{Eq9}
|T^{\theta}_{\theta}|=|T^{\phi}_{\phi}|\leq|T^{t}_{t}|\geq|T^{r}_{r}|\
\end{equation}
between the components of the energy-momentum tensors of physically
acceptable systems.

The action (\ref{Eq3}) also yields the compact radial differential
equation \cite{BekMay}
\begin{equation}\label{Eq10}
\partial_{\alpha}\partial^{\alpha}\psi-\xi R\psi=0\
\end{equation}
for the nonminimally coupled massless scalar field. Substituting
into (\ref{Eq10}) the spherically symmetric curved line element
(\ref{Eq1}), one obtains the characteristic differential equation
\begin{equation}\label{Eq11}
\psi{''}+{1\over2}\big({{4}\over{r}}+\nu{'}-\lambda{'}\big)\psi{'}-\xi
R e^{\lambda}\psi=0\  ,
\end{equation}
which determines the spatial behavior of the static nonminimally
coupled massless scalar field. Following \cite{Hodns,Bha}, we shall
assume that the non-linear massless scalar fields vanish on the
surface $r=R_{\text{s}}$ of the compact reflecting star:
\begin{equation}\label{Eq12}
\psi(r=R_{\text{s}})=0\  .
\end{equation}
In addition, as pointed out in \cite{BekMay}, for physically
acceptable systems the effective gravitational constant,
$G_{\text{eff}}=G(1-8\pi G\xi\psi^2)$ \cite{BekMay}, must be finite
and positive at asymptotic spatial infinity \cite{BekMay}. This physical
requirement yields the asymptotic relations
\begin{equation}\label{Eq13}
-\infty<8\pi\xi\psi^2<1\ \ \ \ \text{for}\ \ \ \ r\to\infty\
\end{equation}
for the radial eigenfunction of the nonminimally coupled scalar
field.

\section{The non-existence theorem for the composed
reflecting-star-nonminimally-coupled-massless-scalar-field configurations}

In the present section we shall explicitly prove that spherically
symmetric neutral reflecting stars {\it cannot} support non-linear
matter configurations made of static {\it nonminimally} coupled
massless scalar fields.

Using the Einstein relation $R=-8\pi T$ between the scalar curvature
and the trace of the energy-momentum tensor, one obtains from Eqs.
(\ref{Eq4}), (\ref{Eq5}), and (\ref{Eq6}) the functional expression
\begin{equation}\label{Eq14}
R=-{{8\pi
e^{-\lambda}}\over{1-8\pi\xi\psi^2}}\Big[\xi\big({{12}\over{r}}+3\nu{'}-3\lambda{'}\big)\psi\psi{'}+
6\xi\psi\psi{''}+(6\xi-1)(\psi{'})^2\Big]\
\end{equation}
for the scalar curvature of the spherically symmetric static
spacetime. Substituting (\ref{Eq14}) into the radial equation
(\ref{Eq11}), one obtains the (rather cumbersome) differential
equation
\begin{equation}\label{Eq15}
\psi{''}\cdot\Big(1+{{48\pi\xi^2\psi^2}\over{1-8\pi\xi\psi^2}}\Big)
+\psi{'}\cdot\Big\{{{2}\over{r}}\Big(1+{{48\pi\xi^2\psi^2}\over{1-8\pi\xi\psi^2}}\Big)+
{1\over2}(\nu{'}-\lambda{'})+{{8\pi\xi\psi}
\over{1-8\pi\xi\psi^2}}\big[3\xi(\nu{'}-\lambda{'})\psi+(6\xi-1)\psi{'}\big]\Big\}=0\
,
\end{equation}
which determines the spatial behavior of the nonminimally coupled
massless scalar field configurations in the curved spacetime.

We shall first prove that the scalar eigenfunction $\psi$ must
approach zero at spatial infinity. From Eqs. (\ref{Eq2}),
(\ref{Eq6}), (\ref{Eq8}), and (\ref{Eq9}), one deduces that
$\psi\psi{'}$ must approach zero asymptotically faster than $1/r^2$.
That is,
\begin{equation}\label{Eq16}
r^2\psi\psi{'}\to0\ \ \ \ \text{for}\ \ \ r\to\infty\ .
\end{equation}
Taking cognizance of the relations (\ref{Eq2}) and (\ref{Eq16}), one
finds that, in the asymptotic far region $M/r\ll 1$, the scalar
radial equation (\ref{Eq15}) can be approximated by the differential
equation \cite{Noteba,Noteco}
\begin{equation}\label{Eq17}
\psi{''}+{{2}\over{r}}\psi{'}=0\ \ \ \ \text{for}\ \ \ \ r\gg M\ ,
\end{equation}
whose general mathematical solution is given by
\begin{equation}\label{Eq18}
\psi(r)={{A}\over{r}}+B\ \ \ \ \text{for}\ \ \ \ r\gg M\ ,
\end{equation}
where $\{A,B\}$ are constants.

One immediately realizes that the scalar function (\ref{Eq18}) with
$B\neq0$ violates the characteristic relation (\ref{Eq16}) for
asymptotically flat spacetimes \cite{Hodpb}. We therefore conclude
that physically acceptable (finite mass) configurations of the
static scalar field, {\it if} they exist, must be characterized by
an asymptotically vanishing radial eigenfunction [see Eq.
(\ref{Eq18}) with $B=0$]
\begin{equation}\label{Eq19}
\psi(r)={{A}\over{r}}\ \ \ \ \text{for}\ \ \ \ r\gg M\ .
\end{equation}

Taking cognizance of the inner boundary condition (\ref{Eq12}) of
the scalar eigenfunction at the surface of the compact reflecting
star, together with the characteristic asymptotic behavior
(\ref{Eq19}) at spatial infinity, one deduces that the
characteristic scalar eigenfunction $\psi$ of the nonminimally
coupled massless scalar field must have (at least) one extremum
point, $r=r_{\text{peak}}$, within the interval
$r_{\text{peak}}\in(R_{\text{s}},\infty)$. In particular, the radial
scalar eigenfunction is characterized by the simple functional
relations
\begin{equation}\label{Eq20}
\{\psi\neq0\ \ \ ; \ \ \ \psi{'}=0\ \ \ ; \ \ \
\psi\cdot\psi{''}<0\}\ \ \ \ \text{for}\ \ \ \ r=r_{\text{peak}}\
\end{equation}
at this extremum point. In addition,
\begin{equation}\label{Eq21}
\psi\cdot\psi{'}\geq0\ \ \ \ \text{for}\ \ \ \
r\in[R_{\text{s}},r_{\text{peak}}]\  .
\end{equation}

\subsection{The no-scalar-hair theorem for generic inner boundary conditions in
the physical regimes $\xi<0$ and $\xi>{1\over 6}$}

Interestingly, as we shall now show explicitly, one can rule out the
existence of asymptotically flat non-linear static configurations
(`hair') made of nonminimally coupled massless scalar fields in the
physical regimes $\xi<0$ and $\xi>1/6$ {\it without} using the inner
boundary condition (\ref{Eq12}) at the surface of the central
compact object.

Substituting (\ref{Eq19}) into (\ref{Eq4}) and (\ref{Eq5}) and using
the asymptotic far region relation (\ref{Eq2}), one finds the
compact expressions
\begin{equation}\label{Eq22}
T^{t}_{t}=(2\xi-1/2){{\psi^2}\over{r^2}}\cdot[1+O(M/r)]\
\end{equation}
and
\begin{equation}\label{Eq23}
T^{r}_{r}=(1/2-4\xi){{\psi^2}\over{r^2}}\cdot[1+O(M/r)]\
\end{equation}
for the components of the energy-momentum tensor which characterizes
the nonminimally coupled massless scalar fields. Interestingly, the
functional expressions (\ref{Eq22}) and (\ref{Eq23}) for the
energy-momentum components yield the far-region relation
\begin{equation}\label{Eq24}
|T^{r}_{r}|>|T^{t}_{t}|\ \ \ \ \text{for}\ \ \ \ \xi<0 \ \
\text{or}\ \ \xi>1/6\  .
\end{equation}

From the inequality (\ref{Eq24}) one immediately reveals the fact
that spatially regular non-linear configurations of the nonminimally
coupled massless scalar fields {\it violate} the characteristic
relation (\ref{Eq9}) imposed by causality on the energy-momentum
components of physically acceptable systems. Our analysis therefore
rules out the existence of spherically symmetric static
configurations made of massless scalar fields nonminimally coupled
to gravity in the physical regimes $\xi<0$ and $\xi>1/6$.

Before we proceed, it is worth emphasizing again that the
no-scalar-hair theorem presented in this subsection is based on the
characteristic asymptotic ({\it far}-region $r\gg M$) behavior of
the nonminimally coupled massless scalar fields. Our theorem, which
excludes nonminimally coupled massless scalar hair in the physical
regimes $\xi<0$ and $\xi>1/6$, is therefore valid for both
black-hole spacetimes with inner {\it attractive} boundary
conditions (at the absorbing horizon of a classical black-hole
spacetime) and for horizonless curved spacetimes with inner {\it
repulsive} boundary conditions (at the surface of a compact
reflecting star).

\subsection{The no-scalar-hair theorem for spherically symmetric neutral reflecting
stars with generic values of the nonminimal coupling parameter
$\xi$}

We start our second no-scalar-hair theorem, which would be valid for
spherically symmetric reflecting stars with {\it generic} values of
the physical coupling parameter $\xi$, by showing that the radial
function
\begin{equation}\label{Eq25}
\Omega(r;\xi)\equiv 1-8\pi\xi\psi^2\
\end{equation}
is positive definite in the interval
$[R_{\text{s}},r_{\text{peak}}]$. Obviously, $\Omega>0$ for $\xi<0$.
As we shall now prove explicitly, $\Omega>0$ also in the $\xi>0$
case. We first note that, taking cognizance of the inner boundary
condition (\ref{Eq12}) at the surface of the compact reflecting
star, one finds the simple relation
\begin{equation}\label{Eq26}
\Omega(r=R_{\text{s}})=1\  .
\end{equation}
Now, suppose that $\Omega$ vanishes at some point $r=r_0$ in the
interval $[R_{\text{s}},r_{\text{peak}}]$. Then, in the vicinity of
$r_0$, one can expand $\Omega(r)$ in the form
\begin{equation}\label{Eq27}
\Omega(r)=\alpha(r-r_0)^{\beta}+O[(r-r_0)^{\gamma}]\ \ \ \ ; \ \ \ \
\gamma>\beta>0\ ,
\end{equation}
which implies [see Eq. (\ref{Eq25})]
\begin{equation}\label{Eq28}
\psi^2(r)=(8\pi\xi)^{-1}\cdot\{1-\alpha(r-r_0)^{\beta}+O[(r-r_0)^{\gamma}]\}\
.
\end{equation}
From (\ref{Eq28}) one finds the leading order behavior
\begin{equation}\label{Eq29}
\psi\psi{'}(r)=-{{\alpha\beta}\over{16\pi\xi}}\cdot(r-r_0)^{\beta-1}\
\end{equation}
in the vicinity of $r_0$. Note that $\psi\psi{'}\geq0$ in the
interval $[R_{\text{s}},r_{\text{peak}}]$ [see Eq. (\ref{Eq21})],
which implies that $\beta$ is odd \cite{Notebo} and
\begin{equation}\label{Eq30}
\alpha\cdot\beta<0\  .
\end{equation}

It now proves useful to explore the spatial behavior of the linear
combination [see Eqs. (\ref{Eq4}), (\ref{Eq5}), and (\ref{Eq6})]
\begin{equation}\label{Eq31}
{\cal T}\equiv
T^{t}_{t}+T^{r}_{r}-T^{\phi}_{\phi}={{e^{-\lambda}}\over{\Omega}}
\big[{{6\xi}\over{r}}\psi\psi{'}+{1\over2}(\psi{'})^2\big]\  .
\end{equation}
From Eqs. (\ref{Eq27}), (\ref{Eq28}), and (\ref{Eq29}), one finds
the leading order behavior
\begin{equation}\label{Eq32}
{\cal T}={{e^{-\lambda(r_0)}}\over{\alpha(r-r_0)}}
\Big[-{{3\alpha\beta}\over{8\pi
r_0}}+{{\alpha^2\beta^2}\over{64\pi\xi}}(r-r_0)^{\beta-1}\Big]\
\end{equation}
in the vicinity of $r_0$. Taking cognizance of Eq. (\ref{Eq30}), one
immediately realizes that, for $\xi>0$, the expression inside the
square brackets in (\ref{Eq32}) is positive definite, which implies
the singular behavior
\begin{equation}\label{Eq33}
{\cal T}\to\infty\ \ \ \ \text{for}\ \ \ \ r\to r_0\
\end{equation}
in the vicinity of $r_0$. However, the linear combination ${\cal
T}\equiv T^{t}_{t}+T^{r}_{r}-T^{\phi}_{\phi}$ of the energy-momentum
components must be finite for physically acceptable systems [see Eq.
(\ref{Eq7})] \cite{BekMay}. One therefore deduces that $\Omega(r)$
{\it cannot} switch signs in the interval
$[R_{\text{s}},r_{\text{peak}}]$. In particular, taking cognizance
of Eq. (\ref{Eq26}), one finds \cite{Noteonz}
\begin{equation}\label{Eq34}
\Omega(r)\equiv 1-8\pi\xi\psi^2>0\ \ \ \ \text{for}\ \ \ \
r\in[R_{\text{s}},r_{\text{peak}}]\  .
\end{equation}

Finally, taking cognizance of the characteristic functional
relations (\ref{Eq20}) and (\ref{Eq34}), one deduces that, at the
extremum point $r=r_{\text{peak}}$ of the scalar field
eigenfunction, the first term on the l.h.s of (\ref{Eq15}) is {\it
non}-zero whereas the second term on the l.h.s of (\ref{Eq15}) is
{\it zero}. Thus, the equality sign in the radial scalar equation
(\ref{Eq15}) {\it cannot} be respected at the extremum point
$r=r_{\text{peak}}$ of the scalar eigenfunction. We therefore
conclude that, for {\it generic} values of the dimensionless
physical parameter $\xi$, the spherically symmetric neutral
reflecting stars {\it cannot} support massless scalar fields
nonminimally coupled to gravity.

\section{Summary}

It is by now well established that spherically symmetric
asymptotically flat black-hole spacetimes with absorbing event
horizons cannot support spatially regular static configurations made
of massless scalar fields nonminimally coupled to gravity
\cite{BekMay,Bek20,Notesa,Hodn}.

It has recently been proved that {\it horizonless} spacetimes
describing compact reflecting stars may share the interesting
no-hair property with the more familiar classical black-hole
spacetimes \cite{Hodns,Bha}. In particular, it has been proved that
compact objects with reflecting (rather than absorbing) boundary
conditions cannot support spatially regular matter configurations
made of minimally coupled scalar fields, vector fields, and tensor
fields \cite{Hodns,Bha}.

In the present paper we have explored the possibility of extending
the regime of validity of the intriguing no-hair property recently
observed \cite{Hodns,Bha} for horizonless curved spacetimes. To this
end, we have studied analytically the static sector of the
non-linearly coupled Einstein-scalar field equations for spatially
regular {\it nonminimally} coupled massless scalar fields.

Using the causality relations (\ref{Eq9}), which characterize the
energy-momentum components of physically acceptable systems
\cite{BekMay}, we have presented a compact no-scalar-hair theorem
which rules out the existence of non-linear static hairy
configurations made of massless scalar fields nonminimally coupled
to gravity in the physical regimes $\xi<0$ and $\xi>1/6$. It is
worth emphasizing again that this theorem (see Sec. IIIa) is valid
for {\it generic} inner boundary conditions. In particular, the
theorem excludes the existence of both hairy scalar configurations
with attractive inner boundary conditions at the absorbing horizon
of a black-hole spacetime and hairy scalar configurations with
repulsive inner boundary conditions at the compact surface of a
reflecting star.

Finally, we have presented a second no-scalar-hair theorem (see Sec.
IIIb) which explicitly proves that spherically symmetric
asymptotically flat horizonless neutral stars with compact
reflecting surfaces cannot support static massless scalar fields
nonminimally coupled to gravity with {\it generic} values of the
dimensionless physical parameter $\xi$.


\bigskip
\noindent
{\bf ACKNOWLEDGMENTS}
\bigskip

This research is supported by the Carmel Science Foundation. I would
like to thank Yael Oren, Arbel M. Ongo, Ayelet B. Lata, and Alona B.
Tea for helpful discussions.



\begin{thebibliography}{99}
\bibitem{Noteit} It is worth mentioning that, contrary to spherically
symmetric asymptotically flat static black-hole spacetimes, rotating
black holes can support stationary spatially regular massive scalar
field configurations in their exterior spacetime regions
\cite{Hodrc,Herkr}.
\bibitem{Noteunit} We shall use natural units in which $G=c=1$.
\bibitem{Noteba} It is worth emphasizing again that
the physical requirement that the effective gravitational constant,
$G_{\text{eff}}=G(1-8\pi G\xi\psi^2)$ \cite{BekMay}, be finite and
positive at asymptotic infinity \cite{BekMay} implies that, for
physically acceptable spacetimes, $\psi$ cannot blow up at spatial
infinity [see Eq. (\ref{Eq13})].
\bibitem{Noteco} Note that the asymptotic property (\ref{Eq13})
of the physically acceptable \cite{BekMay} nonminimally coupled
scalar eigenfunctions implies the asymptotic far-region $(r\gg M)$
relation $1+48\pi\xi^2\psi^2/(1-8\pi\xi\psi^2)\geq1$.
\bibitem{Notebo} Note that, for even values of $\beta$,
the functional expression (\ref{Eq29}) for $\psi\psi{'}$ would
switch signs at $r=r_0$ in contradiction with the characteristic
relation (\ref{Eq21}).
\bibitem{Noteonz} It is worth noting that, substituting into (\ref{Eq14}) the functional
relations (\ref{Eq20}), which characterize the radial eigenfunction
of the scalar field at its extremum point $r=r_{\text{peak}}$, and
using the fact that $R$ must be finite for physically acceptable
systems [see Eq. (\ref{Eq7})], one immediately deduces that
$\Omega(r=r_{\text{peak}})\neq0$.
\end{thebibliography}
\end{document}